\documentclass[twocolumns,aps,prc]{revtex4}

\usepackage{epsfig}
\usepackage{graphicx}

\begin{document}
\title{Asymptotic behavior of nucleon electromagnetic form 
factors in time-like region}

\author{E. Tomasi-Gustafsson}

\email{etomasi@cea.fr}
\affiliation{\it DAPNIA/SPhN, CEA/Saclay, 91191 Gif-sur-Yvette Cedex,
France }
\author{G. I. Gakh}
\altaffiliation{Permanent address:
\it NSC Kharkov Physical Technical Institute, 61108 Kharkov, Ukraine}
\affiliation{\it DAPNIA/SPhN, CEA/Saclay, 91191 Gif-sur-Yvette Cedex,
France }

\date{\today}
\pacs{25.30.Bf, 13.40.-f, 13.40.Gp}


\begin{abstract}
We study the asymptotic behavior of the ratio of Pauli and Dirac electromagnetic nucleon form factors, $F_2/F_1$, in time-like region for different parametrizations built for the space-like region. We investigate how fast the ratio $F_2/F_1$ approaches the asymptotic limits according to the Phragm\`en-Lindel\"of theorem. We show that the QCD-inspired logarithmic behavior of this ratio results in very far asymptotics, experimentally unachievable. This is also confirmed by the normal component of the nucleon polarization, $P_y$, in $e^++e^-\to N+\bar{N}$ (in collisions of unpolarized leptons), which is a very interesting observable, with respect to this theorem. Finally we observe that the $1/Q$ parametrization of $F_2/F_1$ contradicts this theorem.
\end{abstract}
\maketitle

\section{Introduction}
The asymptotic behavior of the ratio $R=F_2/F_1$ of Pauli and Dirac electromagnetic nucleon form factors (FFs) has recently arised much interest from experimental and theoretical point of view. The last experimental data in space-like (SL) region \cite{Jo00}, about the momentum transfer squared ($q^2=-Q^2) dependence$\footnote{In the following text we will use the notation $t=q^2$ in TL region} of the ratio of the Sachs electric and magnetic FFs, $\mu G_{Ep}(Q^2)/ G_{Mp}(Q^2)$ ($\mu$ is the proton magnetic moment), which has been measured with the polarization transfer method \cite{Re68},  changed the belief that the QCD asymptotic behavior of $F_2/F_1\simeq 1/Q^2$ \cite{Ma73} had already been reached  for $Q^2\ge 2$ GeV$^2$ \cite{Le80}. 

The recent data suggested a different behavior of this ratio: $F_2/F_1\simeq 1/\sqrt{Q^2}$. Such dependence has been justified in framework of different theoretical approaches \cite{Ra03,Ra02,Mi02,Fr96,Ho96,Ca00,Wa01,Bo02}. Another approach, confirming the QCD $1/Q^2$ behavior, discovered the importance of logarithmic corrections, 
$R\simeq ln^2(Q^2/\Lambda^2) /Q^2 $ \cite{Be02}, where $\Lambda$ is the soft scale related to the size of the nucleon. Note that the unexpected behavior of the ratio $\mu G_{Ep}(Q^2)/ G_{Mp}(Q^2)$ was predicted before the experiment took place, by a particular VDM model \cite{Ia73} and also in framework of a soliton model \cite{Ho96}.

The assumption of the analyticity of FFs \cite{Dr61} allows to connect the nucleon FFs in SL and time-like (TL) regions and to study the behavior of the ratio $F_2/F_1$ in TL region. The analyticity of FFs, which has been discussed  for example in Ref. \cite{Ga66}, allows to extend a parametrization of FFs available in one kinematical region to the other kinematical region.

Dispersion relation approaches \cite{Ho76,Du03,Ha04}, which are based essentially on the analytical properties of nucleon electromagnetic FFs, can be considered a powerful tool for the description of the $Q^2$ behavior of FFs in the entire kinematical region.

The VDM model \cite{Ia73}, after appropriate treatment of the $\rho$ contribution, can be also extrapolated from the SL region to the TL region \cite{Ia03,Ia04,Bi04}.

The quark-gluon string model \cite{Ka00}  allowed firstly to find the $Q^2$ dependence of the electromagnetic FFs in TL region, in a definite analytical form, which can be continued in the SL region.

One of the problems concerning FFs of pions and nucleons is the large difference in the absolute values in SL and TL regions. For example, at $q^2$=18 GeV$^2$,  the largest value at which proton TL FFs have been measured \cite{An03}, the corresponding values in TL and SL regions differ by a factor of two. The analyticity of FFs allows to apply the Phragm\`en-Lindel\"of 
theorem \cite{Ti39} which gives a rigorous prescription for the asymptotic behavior of 
analytical functions: 
\begin{equation}
\lim_{t\to -\infty} F^{(SL)}(t) =\lim_{t\to \infty} F^{(TL)}(t).
\label{eq:eq1}
\end{equation}
This means that, asymptotically, FFs have the following constraints: 
\begin{enumerate}
\item The imaginary part of FFs, in TL region, vanishes: $ Im F_i(t)\to 0,$ as $t\to \infty $;
\item The real part of FFs, in TL region, coincides with the 
corresponding value in SL region: $ Re  F_i^{(TL)}(t) [t\to \infty ]= F_i^{(SL)}(t) [t\to -\infty ]$, because FFs are real functions in SL region, due to the hermiticity of the corresponding electromagnetic Hamiltonian.
\end{enumerate}
The existing experimental data violate the 
Phragm\`en-Lindel\"of theorem, even at $t$ values as large as 18 GeV$^2$ \cite{ETG01}. In order to test the two requirements stated above, the knowledge of the differential cross section for $e^++e^-\leftrightarrow p+\bar{p}$ is not sufficient, and polarization phenomena have to be studied also. In this respect, T-odd polarization observables, which are determined by $Im F_1F_2^*$, are especially interesting. The simplest of these observables is the $P_y$ component of the proton polarization in $e^++e^-\to p+\bar{p}$ that in general does not vanish, even in collisions of unpolarized leptons \cite{Du96}, or the asymmetry of leptons produced in $p+\bar{p}\to e^++e^-$, in the collision of unpolarized antiprotons with a transversally polarized proton target (or in the collision of transversally polarized antiprotons on an unpolarized proton target) \cite{Bi93}. 

These observables are especially sensitive to different possible parametrizations of the ratio $R$, suggested by QCD and VDM models. Calculations have been done up to $t\simeq $ 40 GeV$^2$ and show that the $P_y$ component remains large in absolute value \cite{Br04}. For example, QCD inspired parametrizations, which fit reasonably well the data in the SL region, predict $|P_y|\simeq$ 35\% up to $t\simeq $ 40 GeV$^2$. Such behavior has to be considered an indication that the corresponding asymptotics are very far, in agreement with the estimations of the quark-gluon string model \cite{Ka00} and VDM approach \cite{Ia03}.

Note another important property of QCD inspired predictions for nucleon FFs: the corresponding $Im F_i(t)$, $t\ge 4m^2$, $i=1,2$ ($m$ is the nucleon mass), either vanish or have a definite sign in the TL region. The previously quoted parametrizations can not apply in the whole TL region: the asymptotic pQCD behavior follows $F_1(t)\simeq t^{-2}$ and $F_2(t)\simeq t^{-3}$ at large $t$, according to the quark counting rules \cite{Ma73}. The superconvergent conditions:
\begin{equation}
\int_{t_0}^{\infty} Im F_i(t)dt=0,~i=1,2
\label{eq:eq2}
\end{equation}
has to be satisfied, where the lower limit corresponds to $t_0=4m_{\pi}^2$, for isovector FFs, and $t_0=9m_{\pi}^2$ for isoscalar FFs, where $m_{\pi}$ is the pion mass.

This implies that the nonzero QCD-contribution to Eq. (\ref{eq:eq2}) has to be compensated by the corresponding non-perturbative contribution of opposite sign. We can expect that such contribution mainly arises from the special region of $t$: $t_0\le t\le 4m^2$, which is unphysical for the process $e^+ + e^-\leftrightarrow p+\bar{p}$. The contribution from the different vector mesons (with different masses) is expected to be very important here. We can say that the superconvergent condition (\ref{eq:eq2}) can be interpreted as a manifestation of the special duality between pQCD, from one side, and the vector meson contribution, from another side \cite{Gr74}. In principle such duality is similar to the well known Gilman-Bloom duality \cite{Bl71}, concerning the electromagnetic properties of the nucleons in SL region, when the deep inelastic electron nucleon scattering is dual to the excitation of different nucleonic resonances in $e^-+N\to e^-+N^*$. Also one can mention the duality in hadron physics relating the high energy behavior of the amplitudes of hadron-hadron scattering, from one side, to the resonance physics, on the other side.

Returning to the unphysical region, $t_0\le t\le 4m^2$, we recall, for completeness, that another interesting physical effect has to be taken into account here: a specific $\bar{N}N$ bound states, or even gluon states with $J^{PC}=1^{--}$ quantum numbers. And, due to the analyticity of FFs, these effects should appear in the SL region of momentum transfer, and should be correlated with the asymptotic behavior of FFs.

Our main aim here is to discuss the asymptotic behavior of the existing parametrizations for $F_2/F_1$ in TL region, from the point of view of the Phragm\`en-Lindel\"of theorem. In particular we will analyze the behavior of $Im (F_2/F_1)$, its convergence to zero and study more particularly the asymptotic behavior  of the $P_y$-component of the proton polarization in $e^+ +e^-\to \bar{p} +p$, which contains equivalent information. For completeness we will also consider the behavior of the ratio ${\cal R}= |F_2/F_1|_{TL}/|F_2/F_1|_{SL}$
which should converge asymptotically to one, following the Phragm\`en-Lindel\"of theorem.

In order to have a quantitative estimation of the corresponding value of the relevant variable, we will use the following prescription, from Ref. \cite{Ia03}: {\it "A function  $f(z)$ is said to be x\% scaled when its value is x\% of the asymptotic value  $f_{as}(z)$. The value at which this condition is met is the solution of the equation $|f(z)|=x |f_{as}(z)|$ "}. For the cases considered here, this definition translates into the following three equations\footnote{When $f_{as}(z)=0$, we take $|f(z)|=x$.}:
\begin{eqnarray}
{\cal F}&=&|Im (F_2/F_1)|/|Re (F_2/F_1)|=\Delta \label{eq:eqf}\\
|P_y|&=&\Delta~ \label{eq:eqp}\\
{\cal R}&=&|F_2/F_1|_{TL}/|F_2/F_1|_{SL}=1+\Delta
\label{eq:eqr}
\end{eqnarray}
where we will take $\Delta=0.1$  and  $\Delta=0.05$ in order to characterize the deviations from the asymptotic predictions of the Phragm\`en-Lindel\"of theorem.

For this aim, we use the following three different parametrizations, which apply in the SL region:
\begin{equation}
\displaystyle\frac{F_2}{F_1}=\displaystyle\frac{a}{\sqrt{(-t)}},~a=1.25\mbox{~GeV}\mbox{~from~Ref.~\protect\cite{Br04}},
\label{eq:eq5}
\end{equation}
\begin{equation}
\displaystyle\frac{F_2}{F_1}=0.17[\mbox{~GeV}^2] \displaystyle\frac{ln^2(-t/\Lambda^2)}{(-t)}\mbox{~with~} \Lambda=0.3\mbox{~GeV} \mbox{~from~Ref.~ \protect\cite{Be02}}, 
\label{eq:eq6}
\end{equation}
and the VDM inspired  parametrization from Ref. \protect\cite{Ia04}:
\begin{equation}
\displaystyle\frac{F_2}{F_1}=\displaystyle\frac{F_2^{(S)}+F_2^{(V)}}{F_1^{(S)}+F_1^{(V)}}
\label{eq:eq7}
\end{equation}
where
\begin{eqnarray*}
F_1^{(S)}(Q^2)&=&
\displaystyle\frac{g(Q^2)}{2}
\left[(1-\beta_\omega-\beta_\phi)+\beta_\omega\displaystyle\frac{\mu_\omega^2}{\mu_\omega^2+Q^2}+\beta_\phi
\displaystyle\frac{\mu_\phi^2}{\mu_\phi^2+Q^2}\right],\\
F_1^{(V)}(Q^2)&=&\displaystyle\frac{g(Q^2)}{2}
\left[(1-\beta_\rho)+\beta_\rho
\displaystyle\frac{\mu_\rho^2+8\Gamma_\rho\mu_\pi/\pi}
{(\mu_\rho^2+Q^2)+(4\mu_\pi^2+Q^2)\Gamma_\rho\alpha(Q^2)/\mu_\pi}\right],\\
F_2^{(S)}(Q^2)&=&
\displaystyle\frac{g(Q^2)}{2}
\left[(\mu_p+\mu_n-1-\alpha_\phi)
\displaystyle\frac{\mu_\omega^2}
{\mu_\omega^2+Q^2}+\alpha_\phi\displaystyle\frac{\mu_\phi^2}{\mu_\phi^2+Q^2}
\right],\\
F_2^{(V)}(Q^2)&=&\displaystyle\frac{g(Q^2)}{2}
\left[(\mu_p-\mu_n-1)
\displaystyle\frac{\mu_\rho^2+8\Gamma_\rho\mu_\pi/\pi}{(\mu_\rho^2+Q^2)+
(4\mu_\pi^2+Q^2)
\Gamma_\rho\alpha(Q^2)/\mu_\pi}\right],
\end{eqnarray*}
where $g(Q^2)=\displaystyle\frac{1}{(1+\gamma Q^2)^2}$ 
and $\alpha(Q^2)=\displaystyle\frac{2}{\pi}
\sqrt{\displaystyle\frac{Q^2+4\mu_\pi^2}{Q^2}}
ln\left[\displaystyle\frac{\sqrt{(Q^2+4\mu_\pi^2)}+\sqrt{Q^2}}{2\mu_\pi}\right]$,
 with the standard values of the masses $m=0.939$~GeV, $\mu_{\rho}=0.77$~GeV, $\mu_\omega=0.78$~GeV, $\mu_\phi=1.02$~GeV, $\mu_\pi=0.139$~GeV and the $\rho$ width $\Gamma_\rho=0.112$~GeV. $\mu_p$ and $\mu_n$ are the magnetic moments of proton and neutron, respectively, whereas $\gamma=0.25$ GeV$^{-2}$, $\beta_{\rho}=0.672$, $\beta_{\omega}=1.102$, $\beta_{\phi}=0.112$, and $\alpha_{\phi}=-0.052$ are parameters fitted on the data.

This paper is organized as follows. In Section II we analyze the $t$-behavior of the imaginary part of the $F_2/F_1$ ratio for different approaches, and estimate the corresponding value of $t$ for deviations of the order of  $\Delta$ from the expected asymptotic values. Then we give the expressions for the  polarization observables accessible through the reaction $e^++e^-\to p\overline p$ in terms of the ratio $F_2/F_1$ and analyze in particular the $P_y$ component of the proton polarization, which depends on the imaginary part of this ratio (Section III). In Section IV we study how the ratio ${\cal R}$ approaches to one, that is the expected value for the asymptotic regime.

\section{ Imaginary part of the nucleon electromagnetic form factors}

Let us recall here the definition of the Phragm\`en-Lindel\"of theorem, which will be the basis of the following discussion. Following \cite{Ti39}: {\it 
if $f(z)\to a $ as $z\to \infty$ along a straight line, and $f(z)\to b$ as $z\to \infty$ along another straight line, and $f(z)$ is regular and bounded in the angle between, then $a=b$ and $f(z)\to a$ uniformly in the angle"}. For the problem considered here, we identify the variable $z$ with the momentum transfer squared $t$. So one of these straight lines can be chosen along the $x$-axis, in the positive direction (in the complex $z$-plane), i.e., for $t$ values corresponding to the TL region, and the other line with negative $x$ direction, with $t$ in the SL region. Assuming the analyticity of FFs, $F_i(t)$, $i=1,2$, in the upper part of the $z$-plane, we satisfy the necessary conditions for the application of the Phragm\`en-Lindel\"of theorem,  for all nucleon FFs, $F_{1,2}(t)$. More exactly, it holds also for the four independent FFs $F_{1,2}^{(S)}(t)$ and $F_{1,2}^{(V)}(t)$, where the upper indices $(S)$ or $(V)$ correspond to isoscalar or isovector electromagnetic FFs of the nucleon. Note that the analytical properties of $F_i(t)$, $i=1,2$, should be discussed namely for the isoscalar and the isovector FFs, and not for proton and neutron, because the unitarity conditions (which allow to calculate the imaginary part of FFs) have the simplest and most transparent form for the isotopic FFs. More exactly, isoscalar(isovector) FFs are determined by intermediate states with odd(even) number of pions.

So, finally, one can write the following four independent relations:
\begin{equation}
\lim_{t\to +\infty} F_{1,2}^{(S,V)}(t) =\lim_{t\to -\infty} F_{1,2}^{(S,V)}(t) 
\label{eq:eq1a}
\end{equation}
as a consequence of the Phragm\`en-Lindel\"of theorem. 

This theorem has other applications in particle physics, such as, for example, the well known theorem of Pomeranchuk \cite{Lo65}, concerning the asymptotic behavior of the total cross sections for $a+b$ and $\bar a +b$ collisions ($a$ and $b$ any hadrons): $\sigma_T(ab)=\sigma_T(\bar a b)$. However, to be rigorous, the applicability of this theorem to FFs, which seems evident, has not been proved up to now\footnote{In Ref. \protect\cite{Go94} one can read {\it "There is, a priori, no general constraint to ensure that the limit of some observable, such as a form factor, should be the same in every direction in the complex plane".}}.

Unfortunately, this theorem does not allow to indicate the physical value of $t$, starting from which it is working at some level of precision. For this aim one needs some additional dynamical information.

In our considerations about nucleon electromagnetic FFs, such information is contained in the parametrizations of FFs. More precisely, we discuss the ratio $F_2/F_1$ for the proton and use those parametrizations which work well in the SL region, where the available precise experimental data allow to constrain the necessary parameters. It is possible to continue analytically such parametrizations to the TL region, using the following prescription \cite{Br04}:
\begin{equation}
ln(-t)=ln(t)-i\pi,~t> 0.
\label{eq:eq4}
\end{equation}
Evidently, the choice of sign for the imaginary part\footnote{Note that in Ref. \protect\cite{Bi93} another sign has been taken: $ln(-t)=ln(t)+i\pi$, whereas in Ref. \protect\cite{Ba00} the formula $ln(-t)=ln(t)\pm i\pi$ has been applied for the analytical continuation from SL to TL region.}, in Eq. (\ref{eq:eq4}), results in strong physical consequences concerning the calculations of any T-odd polarization observable for $e^+ +e^-\leftrightarrow N+\bar N$.

Let us firstly discuss the $t$-behavior of $Im (F_2/F_1)$ in TL region, using the QCD inspired and VDM parametrizations. Following the Phragm\`en-Lindel\"of theorem, the ratio ${\cal F}=|Im(F_2/F_1)|/|Re(F_2/F_1)|$ should converge to zero as $t\to \infty$. And the value of $t$, corresponding to the solution of the equation: ${\cal F}=\Delta$, $\Delta\ll 1$ characterizes how ${\cal F}$ approaches to zero. 

After analytical continuation in TL region, one can see that parametrization
(\ref{eq:eq5}) gives $R\to \infty$, because it reduces in TL region to:
\begin{equation}
\displaystyle\frac{F_2}{F_1}=i\displaystyle\frac{1.25\mbox{~GeV}}{\sqrt{t}},~t>0 ~ \mbox{\protect\cite{Br04}}.
\label{eq:eq8}
\end{equation}
Such parametrization definitely contradicts the Phragm\`en-Lindel\"of theorem because both form factors can not be real at the same time.

This situation is not changed, after a modification suggested to normalize FFs  at $t=0$ \cite{Br04}:
$$\displaystyle\frac{F_2}{F_1}\to\left [\displaystyle\frac{1}{\kappa_p^2} +\displaystyle\frac{t}{(1.25)^2\mbox{~GeV}^2}\right ]^{-1/2},
$$
where $\kappa_p$ is the proton anomalous magnetic moment.
 
Parametrization (\ref{eq:eq6}) results in the following formula for the relative size of the imaginary to the real part ${\cal F}$: 
\begin{equation}
{\cal F}=\displaystyle\frac{2\pi~ln (t/\Lambda^2)}{ln^2 (t/\Lambda^2)-\pi^2},~t>0, 
\label{eq:eq9}
\end{equation}
which implies ${\cal F}\to 0$, if $t\to \infty$, but very slowly. Quantitatively, the condition ${\cal F}=\Delta$ has two solutions:
\begin{equation}
x_{\pm}=ln\displaystyle\frac{t}{\Lambda^2}=
\displaystyle\frac{\pi}{\Delta}\left (1\pm\sqrt{1+\Delta^2}\right ).
\label{eq:eq10}
\end{equation}
For the $x_+$ solution, which should be considered as the physical solution for the TL region, we obtain: 
$$\sqrt{t}\simeq 10^{13} \mbox{~GeV},\mbox{~for~} \Delta=0.1,$$
which represents a very large energy, not far from the Planck scale, $\sqrt{t}=10^{19}$ GeV. This last value corresponds to a deviation of  6.5\% from the expected asymptotical zero value. 

\begin{figure}
\begin{center}
\includegraphics[width=17cm]{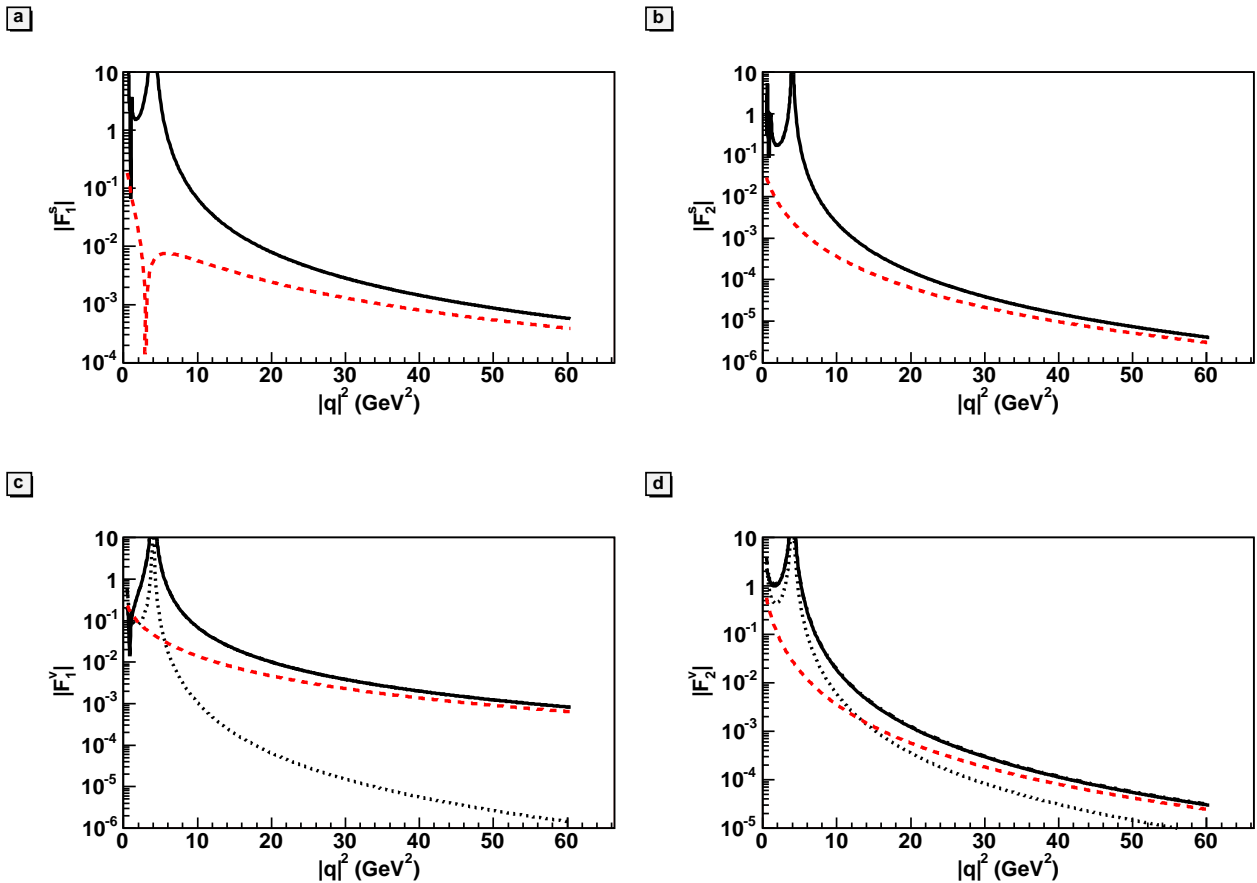}
\caption{\label{Fig:fig1} Isoscalar and isovector FFs in SL and TL regions. 
(a) and (b): $F_1^{(S)}$ and $F_2^{(S)}$ in SL region (dashed line) and in TL region (solid  line), (c) and (d) $F_1^{(V)}$ and $F_2^{(V)}$ in SL region (dashed line) and in TL region: real part (solid line), imaginary part (dotted line) and absolute value (dash-dotted line) which overlaps almost everywhere with the real part.}
\end{center}
\end{figure}

In the model \cite{Ia73}, the isoscalar FFs, $F_{1,2}^{(S)}$, are real in all the kinematical range. Only the isovector FFs, $F_{1,2}^{(V)}$, have non vanishing imaginary part, induced by the $\rho$-meson contribution, which is, however,  one order of magnitude smaller than the real part. The individual FFs are shown in Fig. \ref{Fig:fig1}. A singularity appears in the TL region, for all FFs, due to the dipole term and in $F_1^{(S)}$, due to a compensation of the $\omega$ and $\phi$ contributions.

Taking the parameters from Ref. \cite{Ia73}, one can find:
\begin{equation}
 {\cal F}=\displaystyle\frac{19.36}
{ [1+0.512 ln (\sqrt{t}/m_{\pi})]^2-12.3 ln (\sqrt{t}/m_{\pi})-23.5}
\label{eq:eq11}
\end{equation}
with a faster decreasing, proportional to $[ln(\sqrt{t}/m_{\pi})]^{-2}$, relatively to the previously considered  parametrizations. Such asymptotic behavior leads to $\Delta$=0.1(0.05) for $\sqrt{t}=10^{11}(10^{15})$ GeV, again very far from the region experimentally accessible. Note that the contribution which is linear in logarithm as well as the constant terms in the denominator are important, as they are responsible for the zero of $Re (F_2/F_1)$ at $ln(\sqrt{t}/ m_{\pi})\simeq 45$, which results in a number larger than the asymptotic value.

Recently, the model \cite{Ia73} has been modified with respect to a common factor for all FFs \cite{Ia03,Bi04}:
$$(1-\gamma t)^{-2}\to(1-\gamma e^{i\theta}t)^{-2},$$
where $\theta $ has been taken equal to 53$^0$ and $\gamma$=0.25 GeV$^{-2}$. This term moves the corresponding singularity $t=1/{\gamma}$ to $t= 1/\gamma e^{i\theta}\simeq 4e^{-i\theta}$ GeV$^2$ from the physical region of TL momentum transfer. Such  factor does not modify polarization phenomena as it cancels out. However, such substitution has some shortcomings as it violates the Schwartz reflection symmetry, in the following relation: $F^*(z)=F(z^*)$, and does not satisfy the Phragm\`en-Lindel\"of theorem, because this factor induces: $Im F(t)/Re F(t)\simeq -\tan 2\theta\simeq 3.5$, i.e., a nonzero value in the asymptotic region. 

\section{Polarization observables and asymptotic behavior of the T-odd observable $P_y$}

Let us analyze the polarization observables related to the process $e^++e^-\to p+\bar{p}$ and their asymptotic behavior for the considered parametrizations of $F_2/F_1$. As the $\cos\theta$ dependence is not relevant for the following considerations, for the numerical calculations we take $\theta=45^o$. The $\cos\theta$ dependence is well known in framework of one photon exchange \cite{Du96}, therefore, its measurement can be useful to check the validity of this mechanism at large $Q^2$. It is straightforward to derive the expressions for the polarization observables in terms of $F_2/F_1$ following the formalism   derived in Ref. \cite{Du96}. The reference system is taken as follows: the $z$ axis along the direction of the colliding electron, the $y$ axis normal to the scattering plane, defined by the direction of the electron and of the outgoing proton, and the $x$ axis to form a left-handed coordinate system.

In case of unpolarized beam and target, only a single spin polarization observable does not vanish, the component of the polarization of the scattered proton which is normal to the scattering plane, $P_y$:
\begin{equation}
P_y=-\displaystyle\frac{\tau-1}{\sqrt{\tau}}\displaystyle\frac{Im F_2/F_1}{D},
\label{eq:py}\\
\end{equation}
where $\tau=t/(4m^2)$ and 
$$
D=\displaystyle\frac{3}{2}\left | 1+\displaystyle\frac{F_2}{F_1}\right |^2+
\displaystyle\frac{1}{2\tau}
\left | 1+\tau\displaystyle\frac{F_2}{F_1}\right |^2=
\displaystyle\frac{1}{2}\left [3+8Re \displaystyle\frac{F_2}{F_1}
+\displaystyle\frac{1}{\tau}+(\tau+3)\left |\displaystyle\frac{F_2}{F_1}\right |^2\right ].$$
The double spin coefficients, which do not vanish due to parity and C conservations, are:
\begin{eqnarray}
A_{xx}&=&\displaystyle\frac{1}{2D}\left[1+\displaystyle\frac{1}{\tau}+4Re \displaystyle\frac{F_2}{F_1}+ (1+\tau) \left |\displaystyle\frac{F_2}{F_1}\right |^2\right ],
\label{eq:Axx}\\
A_{yy}&=&\displaystyle\frac{1-\tau}{2\tau D}\left[1-
\tau \left |\displaystyle\frac{F_2}{F_1}\right |^2\right ],
\label{eq:Ayy}\\
A_{xz}&=&\displaystyle\frac{1}{\sqrt{\tau}D}
\left[1+(1+\tau)Re \displaystyle\frac{F_2}{F_1} +\tau \left |\displaystyle\frac{F_2}{F_1}\right |^2\right ],
\label{eq:Axz}\\
A_{zz}&=&\displaystyle\frac{1}{2D}
\left[3-\displaystyle\frac{1}{\tau}+
4Re \displaystyle\frac{F_2}{F_1}
+(3-\tau )\left |\displaystyle\frac{F_2}{F_1}\right |^2\right ],
\label{eq:Azz}
\end{eqnarray}
and they depend on the real part and/or on the modulus of $F_2/F_1$.
The observable $P_y$, which contains the imaginary part of the FFs ratio, can bring information for the comparison of  SL and TL asymptotic behavior.

The following formula for $P_y$, at $\tau\gg 1$, holds for the parametrization (\ref{eq:eq5}):
$$P_y=-\displaystyle\frac{\left(1-\displaystyle\frac{1}{\tau}\right )
\displaystyle\frac{a}{m}}{3+\displaystyle\frac{1}{\tau}+\left (1+\displaystyle\frac{3}{\tau}\right ) \displaystyle\frac{a^2}{4m^2}}
\to P_{y,as}=-\displaystyle\frac{a/m}{3+a^2/(4m^2)}\simeq -0.387.
$$
This parametrization results in non vanishing (negative) asymptotics $P_y$, with large absolute value, in contradiction with the Phragm\`en-Lindel\"of theorem. The behavior of $P_y$ for $1/\tau\ll 1$ can be approximated by:
$$P_y=P_{y,as}\left (1-\displaystyle\frac{p}{\tau}\right ),~
p=1+\displaystyle\frac{3+4m^2/a^2}{1+12m^2/a^2}=1.67.$$
This implies that a 10\%(5\%) difference from asymptotics appears at $
t=58.8(117.6)\mbox{~GeV}^2.$

For the logarithmic parametrization (\ref{eq:eq6}), the asymptotic behavior of $P_y$ is described by:
$$P_y\to -0.19 \displaystyle\frac{ln(t/\Lambda^2)}{\sqrt{\tau}}.$$
One can see that the absolute value decreases with $t$, and one finds $P_y=-10$\%(-5)\% at $t\simeq 350(6000)$ GeV$^2$, still too large to be achieved by experiments.

Finally, the asymptotic behavior of the $P_y$ polarization in the model \cite{Ia73,Ia03} can be described by the following formula:
$$P_y\to \displaystyle\frac{3.5/\sqrt{\tau}}
{\left [1+0.51 ln  (\sqrt{t}/m_{\pi} )
\right ]^2}\to 
\displaystyle\frac
{13.5}{\sqrt{\tau} ln^2  (\sqrt{t}/m_{\pi})} $$
with a faster decreasing with $t$. Note that this polarization is positive in TL region. Moreover the constant term in the denominator is important at large $t$, for example, a value of $P_y$=0.02 is reached at $t=2\cdot 10^6$ GeV$^2$, which corresponds to very far asymptotics.

For the cases discussed above, the large value of $|P_y|$ arises questions about the asymptotic trend of electromagnetic FFs. According to the prescriptions of Phragm\`en-Lindel\"of theorem, $P_y$ should vanish.
 
\section{Difference between the absolute values of $F_2/F_1$ in SL and TL regions}

We mentioned above that the measured values of the magnetic proton FFs are different in TL and SL regions of momentum transfer, up to $t=18$ GeV$^2$, where the TL values of $|G_{Mp}|^2$ exceed by a factor of two the corresponding values in SL region. These values should approach the same number at asymptotic values of $t$. But which number?

Let us analyze the behavior of $|F_2/F_1|$ in TL region, using, again, the considered parametrizations. 
The parametrization (\ref{eq:eq5}) gives $|F_2/F_1|_{SL}=|F_2/F_1|_{TL}$, at any value of $t$. Furthermore, this parametrization gives a specific behavior of the ratio $|G_E|^2/|G_M|^2$ in TL region. One finds:
\begin{equation}
\displaystyle\frac{ |G_E|^2}{|G_M|^2}
=\displaystyle\frac{1+\tau\displaystyle\frac{a^2}{4m^2}}
{1+\displaystyle\frac{a^2}{4m^2\tau}}\to\tau \displaystyle\frac{a^2}{4m^2}=0.44\tau.
\label{eq:eqa}
\end{equation}
Note, in this respect, that up to now the separation of the electric and magnetic contributions to the differential cross section in the TL region has not been realized, yet. The analysis of the experimental data is currently based on two assumptions: either $G_E=0$ or $|G_M|=|G_E|$. The extracted values for $G_M$ according to these prescriptions differ at most by 20\%. 

However, Eq. (\ref{eq:eqa}) suggests another possible relation between $G_E$ and $G_M$, that leads to comparable contributions of the electric and magnetic terms to the cross section, independently on the $t$-value. The resulting value for $G_M$ is 10\% lower than the value corresponding to $|G_M|=|G_E|$, and still does not compensate the observed difference of FFs in SL and TL regions.
  
The parametrization (\ref{eq:eq6}) gives the following relation:
$$
{\cal R}=\displaystyle\frac{|F_2/F_1|_{TL}}{|F_2/F_1|_{SL}}=1 +\displaystyle\frac{ \pi^2}{ln^2(t/\Lambda^2)}.
$$
A deviation of ${\cal R}$ from 1 by 10\%(5\%) is reached at $\sqrt{t}\simeq 43 (337)$ GeV.


\section{Conclusions}
We have analyzed the asymptotic behavior of recently suggested, pQCD inspired,  parametrizations of the ratio of the Dirac and Pauli FFs, $F_2/F_1$. We have based our study on the requirements given by Phragm\`en-Lindel\"of theorem, in particular the equality of FFs in SL and TL regions. As FFs are real in SL region and complex in TL region, this implies that the imaginary part of FFs in TL region vanishes, as well as the polarization of the emitted proton, in the annihilation reaction $e^++e^-\leftrightarrow p+\bar{p}$ (when the colliding particles are unpolarized).

We have shown that the considered parametrizations do not satisfy the asymptotic conditions suggested by the Phragm\`en-Lindel\"of theorem or they do so only for very large values of $Q^2$, well beyond the experimentally accessible range. In particular, the $1/\sqrt{Q^2}$ behavior of this ratio, which reproduces the recent measurements in the SL region, is certainly not compatible with an asymptotic regime, showing that the presently measurable data should be better interpreted in frame of classical nucleon degrees of freedom.

Concerning the double logarithmic parametrization, it has been pointed out long ago \cite{Mi82}, that a suppression to Sudakov type contributions could take place.

The dipole-like formulas for FFs do satisfy the Phragm\`en-Lindel\"of theorem. But such parametrization has the following evident problems:
\begin{itemize}
\item the threshold condition: $G_{EN}(4m^2)=G_{MN}(4m^2)$ is not satisfied,
\item the unitarity conditions for all nucleon FFs are strongly violated, as one should have a branching point at $t=4m_{\pi}^2$ for isovector FFs and  $t=9m_{\pi}^2$ for isoscalar FFs,
\item the prediction in TL region underestimates the experimental data.
\end{itemize}

The analytical continuation of nucleon electromagnetic FFs, presently used to describe the main properties of nucleon structure in SL region of momentum transfer squared (in some models), results as a rule, in an essential imaginary part in TL region. Moreover, the relative value (with respect to the real part) is a very slowly decreasing function of $t$. Such behavior, of course, is in agreement with the Phragm\`en-Lindel\"of theorem, but the corresponding asymptotic regime corresponds to very large values of $t$.
 
The asymptotic regime defined by the prescriptions of the considered models and the asymptotic properties derived from the analyticity of form factors act at a different level. Phragm\`en-Lindel\"of theorem defines the asymptotic conditions without direct connection with QCD. The most evident application of the  Phragm\`en-Lindel\"of theorem in physics is the Pomeranchuk theorem - which relates the asymptotic behavior of the total cross section for $NN$ and $N\bar N$ interaction. This is not QCD regime, because such theorem applies for $t=0$, i.e., to evidently non perturbative physics, despite the fact that the Mandelstam variable $s$ is very large.  So the connection between QCD asymptotics and asymptotics from Phragm\`en-Lindel\"of theorem, from the point of view of hadron FFs is non trivial, as this theorem seems to work for elastic $NN$ and $N\bar N$ amplitudes in the kinematical region where QCD does not apply.

We can consider the present results as an indirect indication of the importance of non perturbative contributions to the physics of the nucleon electromagnetic structure.
\section{Acknowledgments}

The authors acknowledge Prof. M. P. Rekalo for many interesting discussions and ideas, without which this paper would not have been realized in the present form. Thanks are due to Prof. E. Kuraev for his interest to this subject and to C. Duterte for contributing to the earlier stage of this work.

{}

\end{document}